%% file: Main.tex
\newcommand{\CLASSINPUTtoptextmargin}{19mm}%
\newcommand{\CLASSINPUTbottomtextmargin}{43mm}%
\newcommand{\CLASSINPUTinnersidemargin}{12.9mm}%
\newcommand{\CLASSINPUToutersidemargin}{12.9mm}%
\documentclass[conference,10pt,a4paper]{IEEEtran}
\usepackage{amsmath}
\usepackage{times}
\usepackage{graphicx}
\usepackage{multirow}
\usepackage[none]{hyphenat}
\usepackage{float}
\usepackage{subfig}
\usepackage{booktabs}

\usepackage{t1enc}
\usepackage{times}
\usepackage[table]{xcolor}
\usepackage{array}
\usepackage{eso-pic}

\input{EuMW_modify_IEEEtran_18b_CTAN_V4}

\DeclareUnicodeCharacter{24C7}{\textsuperscript{\textregistered}}
\begin{document}
\raggedbottom
%
%
%
\title{An Efficient \textit{W}-/\textit{D}-Band Power Amplifier in a 130~nm SiGe BiCMOS Process}
%
%
\author{%
\IEEEauthorblockN{%
Han~Zhou\textsuperscript{\protect\$}, Yu~Yan\textsuperscript{\protect\#}, Haojie~Chang\textsuperscript{\protect\$}, Herbert~Zirath\textsuperscript{\protect\#}
}
\IEEEauthorblockA{%
\textsuperscript{\protect\$}Faculty of Information Technology and Communication Sciences,
Tampere University, Finland\\
\textsuperscript{\protect\#}Department of Microtechnology and Nanoscience,
Chalmers University of Technology, Sweden\\
han.zhou@tuni.fi
}
}
%
%
\AddToShipoutPictureFG*{%
  \AtPageUpperLeft{%
    \hspace{0.63in}%
    \raisebox{-0.32in}{%
      \parbox{\dimexpr\paperwidth-1.26in\relax}{%
        \centering\footnotesize\itshape
        This is the author-accepted version of a paper accepted for presentation at the
        2026 Asia-Pacific Microwave Conference (APMC 2026).
        \copyright~2026 IEEE. The final published version will be available in IEEE Xplore.
      }%
    }%
  }%
}
\maketitle
%
%
\begin{abstract}
This paper presents a wideband power amplifier (PA) designed and implemented in Infineon Technologies' 130-nm SiGe BiCMOS process for upper $W$-band and lower $D$-band applications. A complete load-pull simulation methodology is carried out, and a band pass filter (BPF)-based matching strategy is employed for the design of the output and inter-stage matching networks. The fabricated PA prototype achieves a small-signal gain 3-dB bandwidth of 71--133~GHz. Moreover, it maintains a relatively flat gain of approximately 15.7~dB over 75--128~GHz, with less than 1-dB fluctuation. The measured saturated output power is 8.8--11.7~dBm, while the measured peak power-added efficiency (PAE) is 7.2--11.1\%. These results demonstrate the potential of SiGe BiCMOS technology for wideband and integrated transmitter front ends operating across the \textit{W}-/\textit{D}-band frequency range.

\end{abstract}

%

\section{Introduction}

The exponential growth of wireless data traffic and the emergence of high-resolution sensing have stimulated significant interest in the upper millimeter-wave (mm-wave) and lower sub-THz spectrum. In particular, the $W$-band and $D$-band offer wide spectral resources and short wavelengths, making them attractive for high-capacity communication links, imaging, radar, and spectroscopy~\cite{Ericsson2022_6GSpectrum, ref2}.

Despite these advantages, transmitter design at $W$-/$D$-band frequencies remains challenging due to large propagation losses, increased passive losses, and the limited power capability of integrated devices at such high frequencies. As a result, the realization of wideband, efficient, and sufficiently powerful power amplifiers (PAs) is essential for practical systems~\cite{intro}. In this context, SiGe BiCMOS is an attractive technology because it enables high-frequency operation while offering a favorable tradeoff among integration capability, speed, and RF performance.

\begin{figure} [t!]
    \centering     
    \includegraphics[width=0.85\columnwidth]{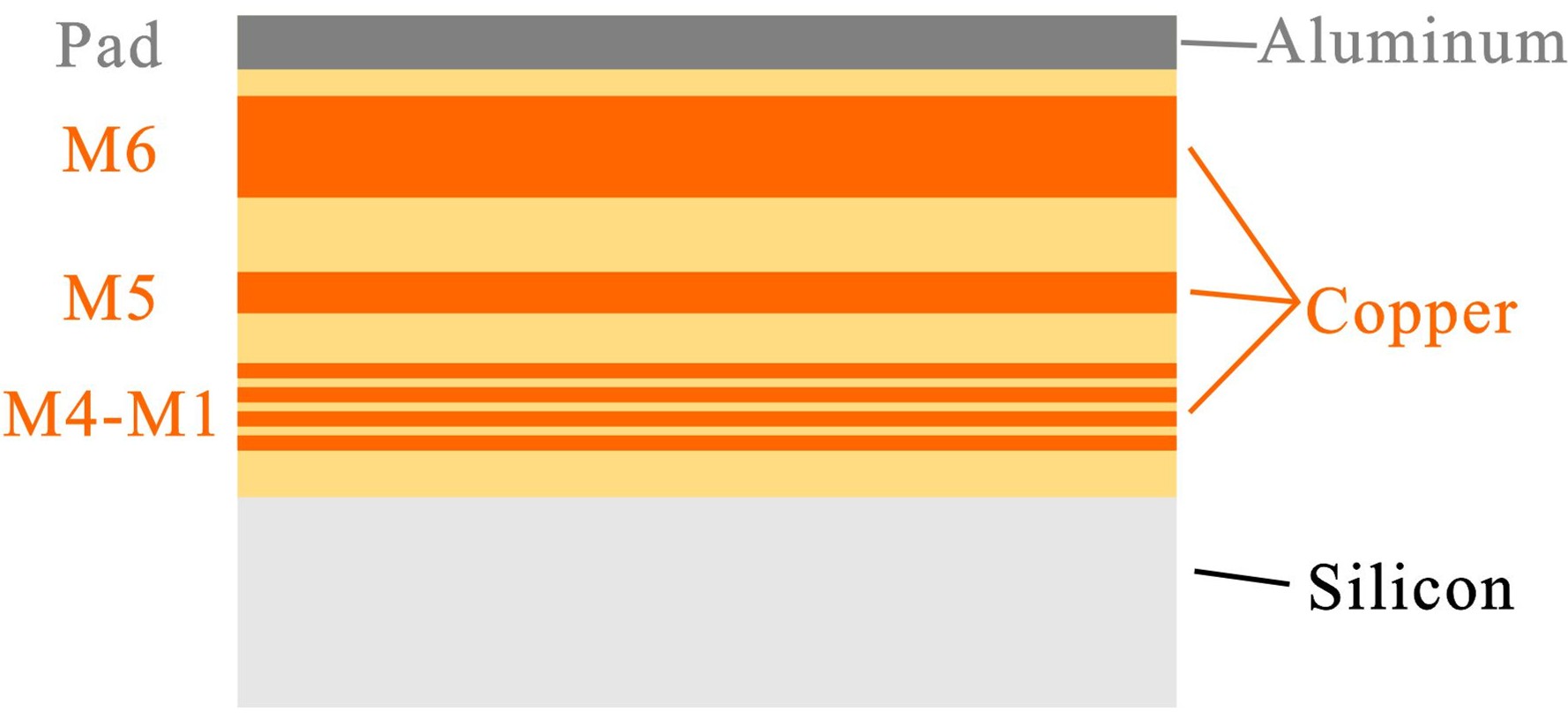}
    \caption{Illustration of the layer stack of the 130nm SiGe BiCMOS (B11HFC) process developed by Infineon Technologies.}
    \label{fig.1}
\end{figure}

This paper presents the design of a PA implemented in Infineon Technologies' 130~nm SiGe BiCMOS process (B11HFC). The technology features high-speed heterojunction bipolar transistors (HBTs) with maximum $f_{\mathrm{t}}$/$f_{\mathrm{max}}$ of 250/370~GHz, making it well suited for upper millimeter-wave integrated circuit design~\cite{SiGeProcess}. As illustrated in Fig.~\ref{fig.1}, the process provides a six-metal-layer stack that supports signal routing and the implementation of high-performance on-chip passive components. Leveraging these technology capabilities, the proposed two-stage cascode PA is designed to achieve wideband operation, high gain, and high efficiency across the 90--130~GHz frequency range, addressing the stringent transmitter requirements of emerging upper millimeter-wave communication and sensing applications.

\begin{figure*} [t!]
    \centering     
    \includegraphics[width=\linewidth]{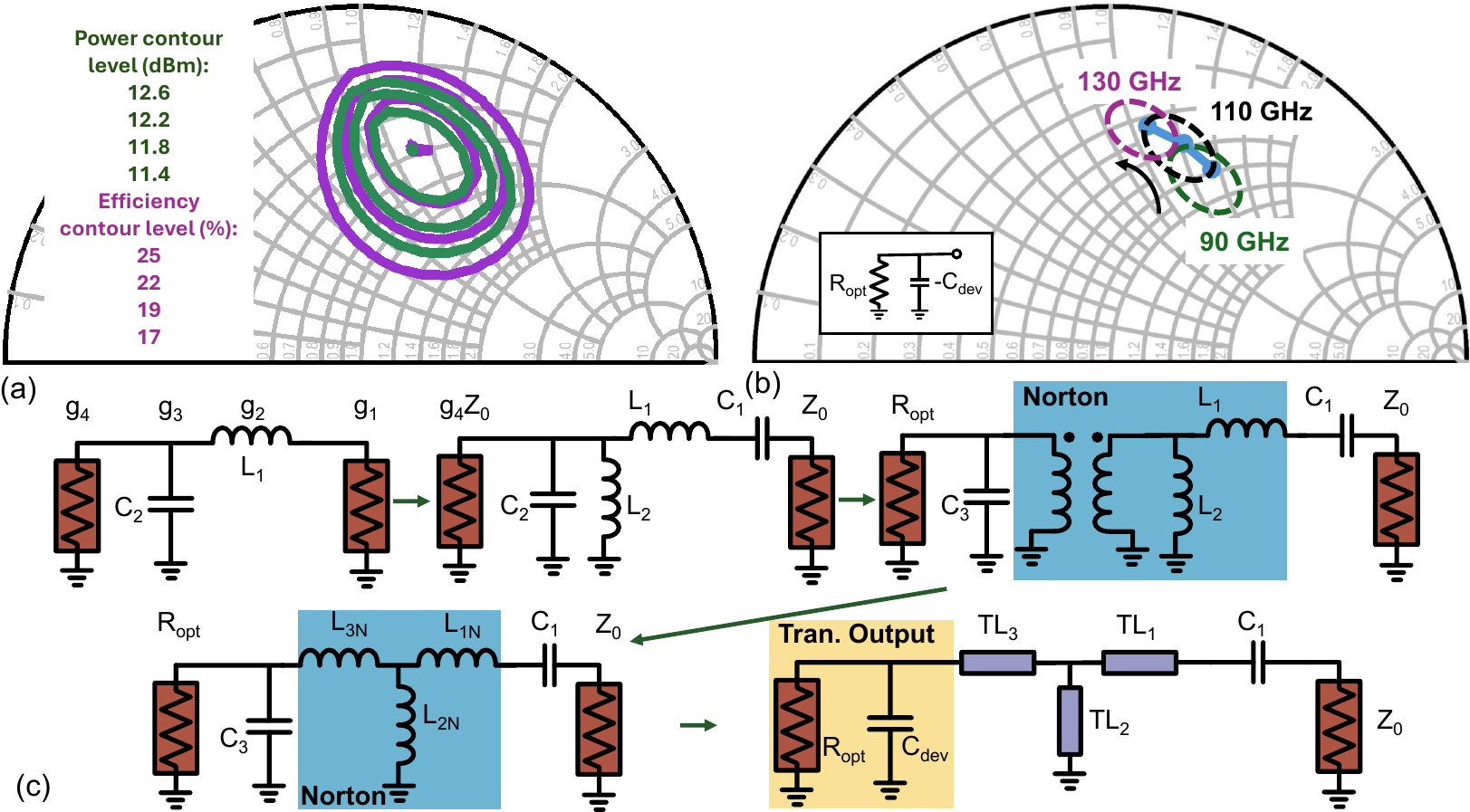}
    \caption{Illustration of (a) load-pull output power and efficiency contours of the cascode PA core at 110~GHz, (b) the $R_{\mathrm{opt}}$-$C_{\mathrm{dev}}$ equivalent output network with acceptable optimum impedance regions at 90, 110, and 130~GHz, and (c) the BPF-based matching network synthesis procedure using the Norton transformation~\cite{BPF}.}
    \label{fig.2}
\end{figure*}

\section{Circuit Design}
As illustrated in Fig.~\ref{fig.2}, the PA design procedure mainly consists of two steps. First, the transistor size is selected to achieve the required output power while maintaining sufficient gain and efficiency. Load-pull simulations are then performed to extract the optimum source and load impedances for each stage~\cite{BBDPAAI}. Second, wideband matching networks are synthesized to transform these optimum impedances to the intended $50~\Omega$ termination at the output ports, or to the input impedance of the subsequent stage for inter-stage matching. To realize low-loss and broadband impedance transformation, a band pass filter (BPF)-based matching approach is adopted.

\subsection{Load-Pull Simulation and Impedance Modeling}
The PA is designed using the high-speed hs1npn HBT device available in the B11HFC process. A transistor size of $10~\mu\mathrm{m}$ is selected to provide sufficient output power while maintaining adequate gain at the target frequencies. To further improve the gain and power delivery capability, a cascode configuration is employed. At higher mm-wave frequencies, however, the cascode topology can suffer from impedance mismatch between the common-emitter and common-base devices, which degrades the deliver power and efficiency. To alleviate this limitation, a compensation transmission line is introduced between the common-emitter and common-base transistors. This inductive compensation improves the inter-device impedance matching and enhances the large-signal performance of the cascode PA~\cite{NoLMPA, RFDAC1, RFDAC2}.

Load-pull simulations are then performed on the cascode PA core to identify the optimum load impedance region. As shown in Fig.~\ref{fig.2}(a), the delivered output power and efficiency contours are plotted on a Smith chart. According to the target output power and efficiency requirements, an acceptable impedance region is selected, which can be approximated by an ellipse. To account for the wideband operation of the PA, load-pull simulations are further carried out across the target frequency range. As shown in Fig.~\ref{fig.2}(b), three impedance ellipses are obtained, representing the optimum load regions at the center frequency and the two band-edge frequencies.

To facilitate the synthesis of the output matching network, the transistor output is modeled by an equivalent parallel network consisting of the optimum resistance $R_{\mathrm{opt}}$ and the device parasitic capacitance $C_{\mathrm{dev}}$~\cite{HW1, 6GPP_Han}. The conjugate impedance presented by the output matching network is then required to lie within the desired load-pull impedance region across the operating frequency range, as illustrated in Fig.~\ref{fig.2}(b). Therefore, the output matching problem can be formulated as a broadband impedance transformation from the parallel $R_{\mathrm{opt}}$-$C_{\mathrm{dev}}$ network to the external $50~\Omega$ load.

\subsection{Band Pass Filter-Based Matching Network Design}
The BPF-based matching network is synthesized from a normalized low-pass prototype~\cite{BPF}. The prototype coefficients $g_i$ are first calculated for a reference system with $1~\Omega$ impedance and a cutoff angular frequency of $1~\mathrm{rad/s}$. The normalized prototype is then transformed into the desired band-pass response by applying frequency scaling with the angular bandwidth $\Delta\omega=2\pi(f_2-f_1)$, where $f_1$ and $f_2$ denote the lower and upper passband edges, respectively. After frequency scaling, impedance scaling is applied to transform the network to the required impedance level $Z_0$. In this step, the series elements are multiplied by $Z_0$, whereas the shunt elements are divided by $Z_0$. By applying the above scaling procedure, the element values of the low-pass prototype, including $L_1=g_1Z_0/\Delta\omega$, $C_2=g_2/(\Delta\omega Z_0)$, and $g_4Z_0$, are obtained. The low-pass network is then converted into a BPF network by resonating each series and shunt element at the geometric mean angular frequency $\omega_c=2\pi\sqrt{f_1f_2}$. Accordingly, the additional resonating elements ($C_1$ and $L_2$), can be derived from the corresponding resonance conditions.

After the BPF network is synthesized, an ideal transformer with an impedance transformation ratio of $n^2$ is introduced to transform the terminal impedance from $g_4Z_0$ to the optimum load resistance $R_{\mathrm{opt}}$. To obtain a physically realizable matching topology, a Norton transformation is applied, replacing the ideal transformer with an equivalent tee-network of inductors, as shown in Fig.~\ref{fig.2}(c)~\cite{BPF}. The device parasitic capacitance $C_{\mathrm{dev}}$ is incorporated into the synthesis procedure, enabling the network to absorb the transistor output capacitance. This approach can also be applied to the inter-stage matching network, since the series capacitance $C_1$ and the resistive component provide a suitable equivalent representation for the input impedance of the subsequent cascaded transistor stage~\cite{input}.

\section{Circuit Implementation}
\begin{figure}[t]
    \centering

    \subfloat[]{
        \includegraphics[width=\linewidth]{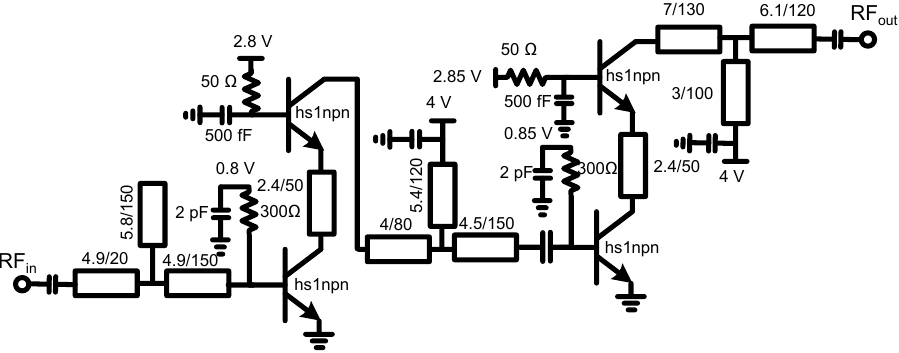}
    }
    \vspace{0.8mm}
    \subfloat[]{
        \includegraphics[width=0.97\linewidth]{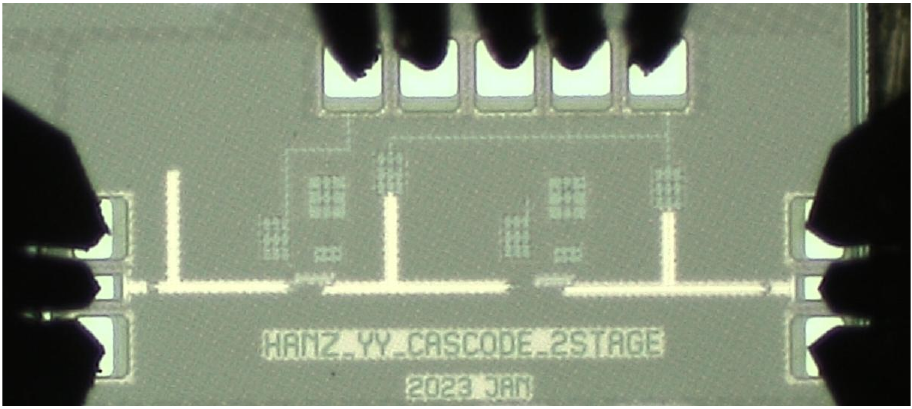}
    }

    \caption{(a) Circuit schematic of the designed PA prototype and (b) chip photograph under a microscope. The chip size is $1030\times490~\mu\mathrm{m}^2$.}
    \label{fig.3}
\end{figure}

Fig.~\ref{fig.3}(a) shows the complete schematic of the proposed two-stage cascode PA. A microphotograph of the fabricated prototype is presented in Fig.~\ref{fig.3}(b). The PA occupies an active area of $0.5~\mathrm{mm^2}$, including the RF and DC probe pads.

To optimize both efficiency and linearity, different bias conditions are adopted for the driver and output stages. The driver stage is biased close to Class-B operation with a base bias voltage of 0.8~V, while the output stage is biased in Class-AB operation with a base bias voltage of 0.85~V. In addition to improving efficiency, this biasing strategy helps mitigate the overall AM-PM distortion of the PA~\cite{AMPM}.

\section{Measurement Results}
\begin{figure}[t]
    \centering

    \subfloat[]{
        \includegraphics[width=\linewidth]{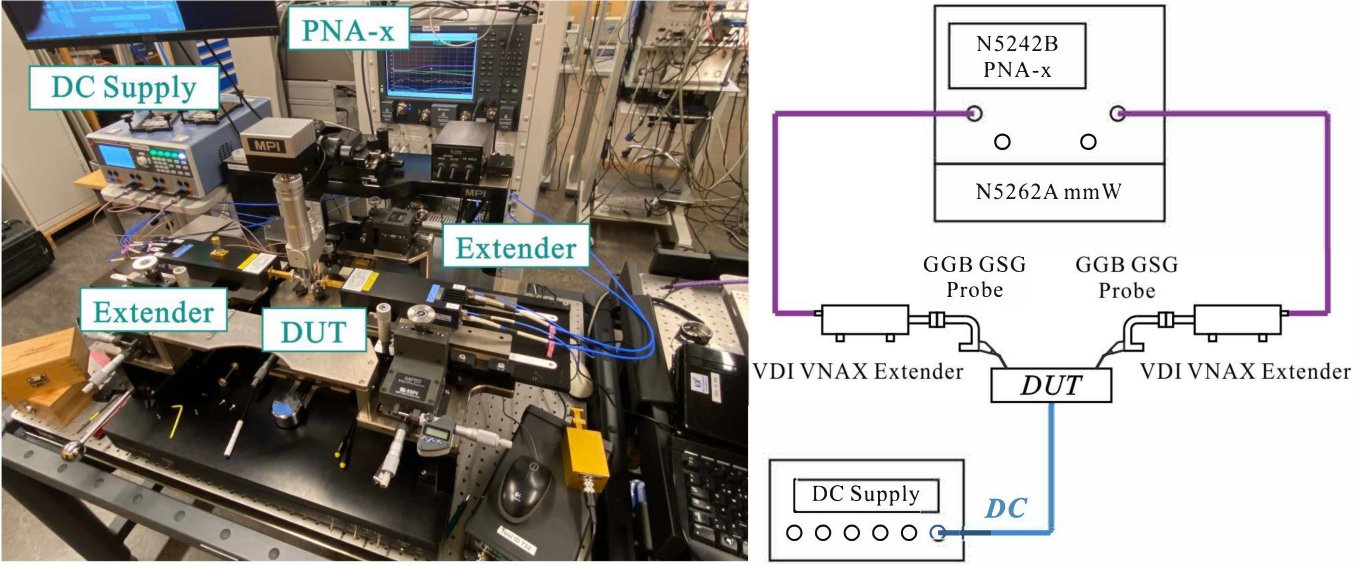}
    }
    \vspace{0.8mm}
    \subfloat[]{
        \includegraphics[width=\linewidth]{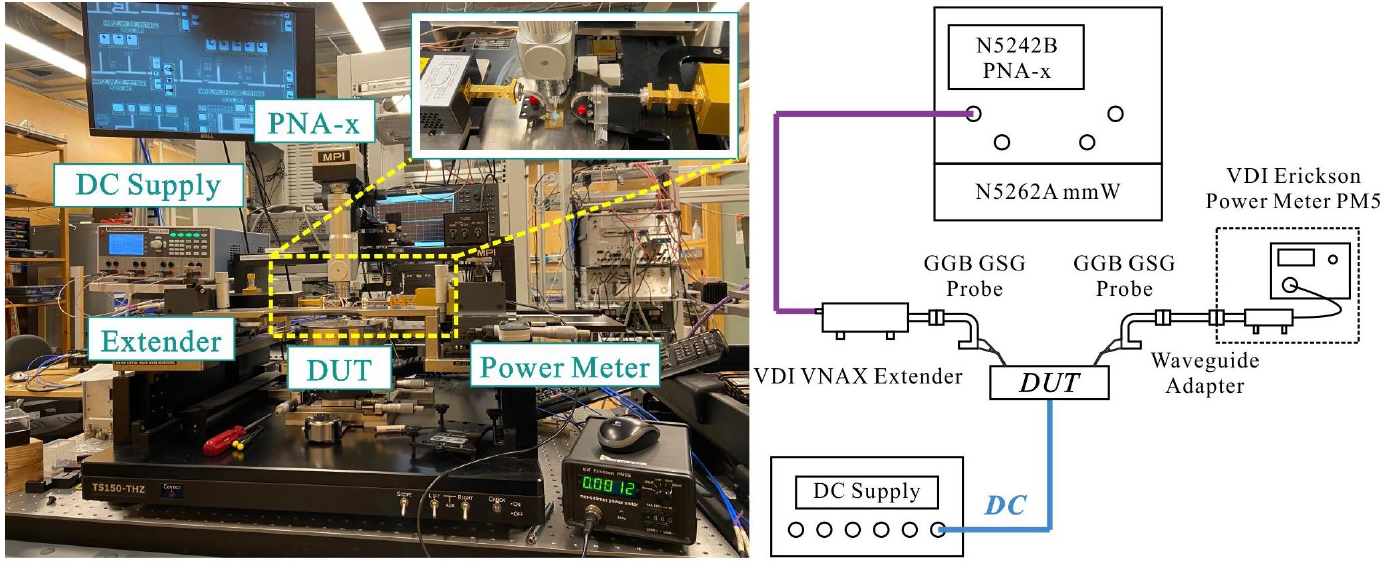}
    }

    \caption{Measurement setups and block diagrams for (a) small-signal characterization and (b) large-signal continuous-wave characterization.}
    \label{fig.4}
\end{figure}

The fabricated prototype is mounted on a copper carrier for characterization. DC and RF probes are used on a probe station to bias and measure the PA. For small-signal characterization, E-band (WR12), F-band (WR8), and D-band (WR6) frequency extenders are employed to cover the entire frequency range of interest. The overlapping frequency coverage of these extenders also enables cross-validation of the measured results. The complete small-signal measurement setup and corresponding block diagram are shown in Fig.~\ref{fig.4}(a). For large-signal characterization, an F-band frequency extender is used for signal generation, while the output power is measured using a VDI Erickson power meter. The large-signal measurement setup is shown in Fig.~\ref{fig.4}(b).

%



\begin{figure} [t!]
    \centering     
    \includegraphics[width=0.9\columnwidth]{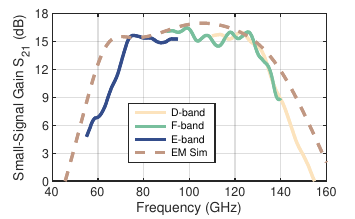}
    \caption{Measured (solid lines, using E-, F-, and D-band extenders) and simulated (dashed lines) small-signal gain of the fabricated PA prototype.}
    \label{fig.5}
\end{figure}

\begin{figure} [t!]
    \centering     
    \includegraphics[width=0.9\columnwidth]{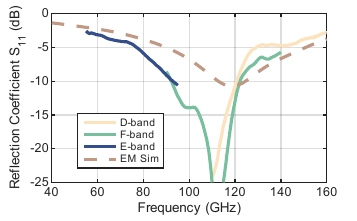}
    \caption{Measured (solid lines, using E-, F-, and D-band extenders) and simulated (dashed lines) input return loss of the fabricated PA prototype.}
    \label{fig.6}
\end{figure}

\begin{figure} [t!]
    \centering     
    \includegraphics[width=0.9\columnwidth]{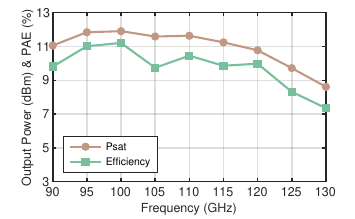}
    \caption{Measured delivered output power and peak PAE versus frequency.}
    \label{fig.7}
\end{figure}

As shown in Fig.~\ref{fig.5} and Fig.~\ref{fig.6}, the measured S-parameters of the fabricated prototype exhibit relatively good agreement with the EM-simulated results, with only minor deviations observed at the band edges, which can be attributed to process variations and device modeling inaccuracies. The measured 3-dB small-signal bandwidth spans 71--133~GHz, demonstrating the wideband capability of the proposed design. Furthermore, the prototype maintains a relatively flat gain of approximately 15.7~dB over the 75--128~GHz frequency range, with less than 1~dB gain variation. Fig.~\ref{fig.7} presents the large-signal measurement results. The PA achieves a saturated output power of 8.8--11.7~dBm, while the peak power-added efficiency (PAE) reaches 7.2--11.1\% across the 90--130~GHz frequency range. 

\begin{table}[t!]
    \centering
    \caption{Summary Of State-of-the-Art \textit{W}-/\textit{D}-Band PAs.}
    \begin{tabular}{ c c c c c c }
    \toprule  
    \multirow{2}{*}{ Ref.} & \multirow{2}{*}{Technology} & BW & Gain & PAE  & P\textsubscript{SAT} \\
    & & (GHz) & ($\%$) & ($\%$) & (dBm) \\

    \midrule
    \multirow{1}{*}{ \cite{tb1}'25} & 130nm SiGe & \multirow{1}{*}{84--110} & 14.2 & 3.1 &  10.2  \\
    \midrule
    \multirow{1}{*}{ \cite{tb2}'25} & 130nm SiGe & \multirow{1}{*}{120--145} & 20.0& 7.0 & 15.0  \\ 
    \midrule
    \multirow{1}{*}{ \cite{tb3}'25} & 45nm CMOS & \multirow{1}{*}{117--132} & 15.5 & 14.6 &  11.9  \\
    \midrule
    \multirow{1}{*}{ \cite{tb4}'22} & 250nm InP& \multirow{1}{*}{133--147} & 8.4 & 22.5 &  17.3 \\
 
    \midrule
    \multirow{1}{*}{\textbf{This Work}} & \textbf{130nm SiGe} & \textbf{71--133} & \textbf{16.7} & \textbf{11.1} & \textbf{11.7} \\
 
    \bottomrule    
    \end{tabular}\\
    \vspace{1mm}
\footnotesize{$^{a}$ BW stands for 3-dB small-signal bandwidth. $^{b}$ Gain, PAE, and $P_{\mathrm{SAT}}$ refer to the highest in-band measured values.}
    \label{tab.1}
\end{table}

As demonstrated in Table~\ref{tab.1}, recently reported \textit{W}-~and \textit{D}-band PAs are summarized and compared with the proposed design. It can be clearly seen that the proposed PA achieves a highly competitive bandwidth and high gain despite its only two-stage cascaded topology. In addition, it demonstrates relatively good PAE and output delivered power considering its single-way configuration.

\section{Conclusion}

This paper presents a two-stage \textit{W}- and \textit{D}-band cascode PA implemented in Infineon Technologies' 130-nm SiGe BiCMOS process. A comprehensive load-pull-based simulation methodology is employed, and a BPF-based matching strategy is adopted for the design of the output and inter-stage matching networks. The fabricated prototype achieves a peak small-signal gain of 16.7~dB, with 3-dB and 1-dB gain bandwidths of 71--133~GHz and 75--128~GHz, respectively. Large-signal measurements further validate the performance of the PA, demonstrating a peak saturated output power of 11.7~dBm and a peak PAE of 11.1\%.
\section*{Acknowledgment}
The authors would like to thank Infineon Technologies for the fabrication of the chips.


\bibliographystyle{IEEEtran}

\bibliography{IEEEabrv,mybibfile}

\end{document}

%% file: EuMW_modify_IEEEtran_18b_CTAN_V4.tex
\makeatletter

\def\@maketitle{\newpage
\bgroup\par\addvspace{0.5\baselineskip}\centering%
\ifCLASSOPTIONtechnote
   {\bfseries\large\@IEEEcompsoconly{\sffamily}\@title\par}\vskip 1.3em{\lineskip .5em\@IEEEcompsoconly{\sffamily}\@author
   \@IEEEspecialpapernotice\par{\@IEEEcompsoconly{\vskip 1.5em\relax
   \@IEEEtitleabstractindextextbox{\@IEEEtitleabstractindextext}\par
   \hfill\@IEEEcompsocdiamondline\hfill\hbox{}\par}}}\relax
\else
   \vskip0.2em{\EuMWtitlesize\ifCLASSOPTIONtransmag\bfseries\LARGE\fi\@IEEEcompsoconly{\sffamily}\@IEEEcompsocconfonly{\normalfont\normalsize\vskip 2\@IEEEnormalsizeunitybaselineskip
   \bfseries\Large}\@title\par}\vskip1.0em\par
   \ifCLASSOPTIONconference%
      {\@IEEEspecialpapernotice\mbox{}\vskip\@IEEEauthorblockconfadjspace%
       \mbox{}\hfill\begin{@IEEEauthorhalign}\@author\end{@IEEEauthorhalign}\hfill\mbox{}\par}\relax
   \else
      \ifCLASSOPTIONpeerreviewca
         {\@IEEEcompsoconly{\sffamily}\@IEEEspecialpapernotice\mbox{}\vskip\@IEEEauthorblockconfadjspace%
          \mbox{}\hfill\begin{@IEEEauthorhalign}\@author\end{@IEEEauthorhalign}\hfill\mbox{}\par
          {\@IEEEcompsoconly{\vskip 1.5em\relax
           \@IEEEtitleabstractindextextbox{\@IEEEtitleabstractindextext}\par\hfill
           \@IEEEcompsocdiamondline\hfill\hbox{}\par}}}\relax
      \else
         \ifCLASSOPTIONtransmag
           {\@IEEEspecialpapernotice\mbox{}\vskip\@IEEEauthorblockconfadjspace%
            \mbox{}\hfill\begin{@IEEEauthorhalign}\@author\end{@IEEEauthorhalign}\hfill\mbox{}\par
           {\vspace{0.5\baselineskip}\relax\@IEEEtitleabstractindextextbox{\@IEEEtitleabstractindextext}\vspace{-1\baselineskip}\par}}\relax
         \else
           {\lineskip.5em\@IEEEcompsoconly{\sffamily}\sublargesize\@author\@IEEEspecialpapernotice\par
           {\@IEEEcompsoconly{\vskip 1.5em\relax
            \@IEEEtitleabstractindextextbox{\@IEEEtitleabstractindextext}\par\hfill
            \@IEEEcompsocdiamondline\hfill\hbox{}\par}}}\relax
         \fi
      \fi
   \fi
\fi\par\addvspace{0.0\baselineskip}\egroup}

\def\EuMWtitlesize{\@setfontsize{\EuMWtitlesize}{24}{24pt}}
\def\EuMWauthorsize{\@setfontsize{\EuMWauthorsize}{11}{11pt}}
\def\EuMWaffilsize{\@setfontsize{\EuMWaffilsize}{10}{10pt}}
\def\EuMWcaptionsize{\@setfontsize{\EuMWcaptionsize}{9}{10pt}}
\def\EuMWbibsize{\@setfontsize{\EuMWbibsize}{8}{10pt}}

\def\@IEEEauthorblockNstyle{\EuMWauthorsize\@IEEEcompsocnotconfonly{\sffamily}\@IEEEcompsocconfonly{\large}}
\def\@IEEEauthorblockAstyle{\EuMWaffilsize\@IEEEcompsocnotconfonly{\sffamily}\@IEEEcompsocconfonly{\itshape}\@IEEEcompsocconfonly{\large}}
\def\@IEEEauthordefaulttextstyle{\EuMWauthorsize\@IEEEcompsocnotconfonly{\sffamily}\sublargesize}

\def\thebibliography#1{\section*{\refname}%
    \addcontentsline{toc}{section}{\refname}%
    \EuMWbibsize\@IEEEcompsocconfonly{\small}\vskip 0.3\baselineskip plus 0.1\baselineskip minus 0.1\baselineskip
    \list{\@biblabel{\@arabic\c@enumiv}}%
    {\settowidth\labelwidth{\@biblabel{#1}}%
    \leftmargin\labelwidth
    \advance\leftmargin\labelsep\relax
    \itemsep \IEEEbibitemsep\relax
    \usecounter{enumiv}%
    \let\p@enumiv\@empty
    \renewcommand\theenumiv{\@arabic\c@enumiv}}%
    \let\@IEEElatexbibitem\bibitem%
    \def\bibitem{\@IEEEbibitemprefix\@IEEElatexbibitem}%
\def\newblock{\hskip .11em plus .33em minus .07em}%
\ifCLASSOPTIONtechnote\sloppy\clubpenalty4000\widowpenalty4000\interlinepenalty100%
\else\sloppy\clubpenalty4000\widowpenalty4000\interlinepenalty500\fi%
    \sfcode`\.=1000\relax}

\long\def\@makecaption#1#2{%
\ifx\@captype\@IEEEtablestring%
\par\@IEEEtabletopskipstrut
\else
\@IEEEfigurecaptionsepspace
\fi
\setbox\@tempboxa\hbox{\normalfont\footnotesize {#1.}\nobreakspace\nobreakspace #2}%
\ifdim \wd\@tempboxa >\hsize%
\setbox\@tempboxa\hbox{\normalfont\footnotesize {#1.}\nobreakspace\nobreakspace}%
\parbox[t]{\hsize}{\normalfont\footnotesize\noindent\unhbox\@tempboxa#2}%
\else
\ifCLASSOPTIONconference \hbox to\hsize{\normalfont\footnotesize\hfil\box\@tempboxa\hfil}%
\else \hbox to\hsize{\normalfont\footnotesize\box\@tempboxa\hfil}%
\fi\fi
\ifx\@captype\@IEEEtablestring%
\@IEEEtablecaptionsepspace
\else
\fi}

\newlength\tablecaptiontotableskip
\newlength\figuretocaptionskip
\def\@IEEEfigurecaptionsepspace{\vskip\figuretocaptionskip\relax}%
\def\@IEEEtablecaptionsepspace{\vskip\tablecaptiontotableskip\relax}%

\def\abstract{\normalfont%
\@IEEEabskeysecsize\bfseries\textit{\abstractname}\,\bfseries\textit{---}\,%
\@IEEEgobbleleadPARNLSP}%

\def\IEEEkeywords{\normalfont%
\@IEEEabskeysecsize\bfseries\textit{\IEEEkeywordsname}\,\bfseries\textit{---}\,%
\@IEEEgobbleleadPARNLSP}%
\def\endIEEEkeywords{\relax\vspace{0.67ex}%
\par\if@twocolumn\else\endquotation\fi%
\normalsize\normalfont}%

\def\@IEEEauthorblockNtopspace{0ex}
\def\@IEEEauthorblockAtopspace{1mm}
\def\tablename{Table}
\def\thetable{\arabic{table}}
\def\IEEEkeywordsname{Keywords}
\def\subsubsection{\@startsection{subsubsection}{3}{\z@}{1.5ex plus 1.5ex minus 0.5ex}%
{0.7ex plus .5ex minus 0ex}{\normalfont\normalsize\itshape}}%
\newlength{\CPheadmatchindent}%
\def\@seccntformat#1{\hbox to\CPheadmatchindent{\csname the#1dis\endcsname}\hskip 0.1em \relax}
\IEEEilabelindentA \parindent
\IEEEilabelindent \IEEEilabelindentA
\IEEEelabelindent \parindent
\IEEEdlabelindent \parindent
\IEEElabelindent \parindent
\makeatother